\documentclass[sigconf]{acmart}
\usepackage{multirow}
\usepackage{colortbl}
\definecolor{lightblue}{HTML}{D1EDFF}

\AtBeginDocument{%
  \providecommand\BibTeX{{%
    \normalfont B\kern-0.5em{\scshape i\kern-0.25em b}\kern-0.8em\TeX}}}

\copyrightyear{2023}
\acmYear{2023}
\setcopyright{acmlicensed}
\acmConference[CIKM '23]{Proceedings of the 32nd
ACM International Conference on Information and Knowledge
Management}{October 21--25, 2023}{Birmingham, United Kingdom}
\acmBooktitle{Proceedings of the 32nd ACM International Conference on
Information and Knowledge Management (CIKM '23), October 21--25, 2023,
Birmingham, United Kingdom}
\acmPrice{15.00}
\acmDOI{10.1145/3583780.3614805}
\acmISBN{979-8-4007-0124-5/23/10}

\newcommand{\ie}{\emph{i.e., }}
\newcommand{\eg}{\emph{e.g., }}

\newcommand{\wrt}{\emph{w.r.t. }}

\newcommand{\aka}{\emph{a.k.a. }}

\begin{document}

\title{CDR: Conservative Doubly Robust Learning for Debiased Recommendation}

\author{Zijie Song}
\email{catshark@zju.edu.cn}
\affiliation{%
  \institution{Zhejiang University}
  \city{HangZhou}
  \state{ZheJiang}
  \country{China}
}

\author{Jiawei Chen}
\authornote{Jiawei Chen is the corresponding author.}
\email{sleepyhunt@zju.edu.cn}
\affiliation{%
  \institution{Zhejiang University}
  \city{HangZhou}
  \state{ZheJiang}
  \country{China}
}

\author{Sheng Zhou}
\email{zhousheng_zju@zju.edu.cn}
\affiliation{%
  \institution{Zhejiang University}
  \city{HangZhou}
  \state{ZheJiang}
  \country{China}
}

\author{Qihao Shi}
\email{shiqihao321@zju.edu.cn}
\affiliation{%
  \institution{Hangzhou City University}
  \city{HangZhou}
  \state{ZheJiang}
  \country{China}
}

\author{Yan Feng}
\email{	fengyan@zju.edu.cn}
\affiliation{%
  \institution{Zhejiang University}
  \city{HangZhou}
  \state{ZheJiang}
  \country{China}
}

\author{Chun Chen}
\email{chenc@cs.zju.edu.cn}
\affiliation{%
  \institution{Zhejiang University}
  \city{HangZhou}
  \state{ZheJiang}
  \country{China}
}

\author{Can Wang}
\email{wcan@zju.edu.cn}
\affiliation{%
  \institution{Zhejiang University}
  \city{HangZhou}
  \state{ZheJiang}
  \country{China}
}

\renewcommand{\shortauthors}{Zijie Song and Jiawei Chen, et al.}

\begin{abstract}
  % In recommendation systems, user behavior is observed rather than experimentally generated, resulting in widespread bias in the data. Debiasing tasks have become a key challenge in recommendation systems. Recently, Doubly Robust method (DR) has become a research hot topic due to its excellent performance and outstanding double robustness properties. However, our experiments show that DR-JL suffers greatly from the impact of poisonous imputation errors, preventing its generalization bound from reaching the theoretical optimum. To address this issue, this paper proposes Conservative DR-JL, which can determine the accuracy of imputation errors and filter out poisonous imputation errors. Theoretical analysis shows that CDR has lower bias and a better generalization bound. Furthermore, our experiments demonstrate that CDR significantly improves performance across multiple baselines and outperforms the best existing algorithms.

   In recommendation systems (RS), user behavior data is observational rather than experimental, resulting in widespread bias in the data. Consequently, tackling bias has emerged as a major challenge in the field of recommendation systems. Recently, Doubly Robust Learning (DR) has gained significant attention due to its remarkable performance and robust properties. However, our experimental findings indicate that existing DR methods are severely impacted by the presence of so-called \textit{Poisonous Imputation}, where the imputation significantly deviates from the truth and becomes counterproductive.

    To address this issue, this work proposes Conservative Doubly Robust strategy (CDR) which filters imputations by scrutinizing their mean and variance. Theoretical analyses show that CDR offers reduced variance and improved tail bounds.
  In addition, our experimental investigations illustrate that CDR significantly enhances performance and can indeed reduce the frequency of poisonous imputation.
\end{abstract}





% \received{3 June 2023}
% \received[revised]{5 August 2023}
% \received[accepted]{18 August 2023}

\maketitle
\allowdisplaybreaks[4]
\section{Introduction}

Enabled by a variety of deep learning techniques, the field of recommendation systems (RS) has seen significant advancements \cite{zhang2020explainable, zhao2019deep}. Nonetheless, the direct application of these advanced RS models in real-world scenarios is often impeded by the presence of numerous biases. Among these, selection bias is especially prominent, referring to the occurrence that the observed data might not faithfully represent the entirety of user-item pairs \cite{marlin2009collaborative, marlin2012collaborative}. Selection bias has detrimental effects not only on the accuracy of the recommendations, but it may also foster unfairness and potentially exacerbate the Matthew effect \cite{zhang2021causal, gao2022cirs, marlin2009collaborative}.

A myriad of methods to counter selection bias have been introduced in recent years. These approaches fall primarily into three categories: 1) Generative models \cite{kim2014bayesian, wang2020information, wang2022invariant}, which resorts to a causal graph to depict the generative process of observed data and infer user true preference accordingly. However, given the complexity of real-world RS scenarios, accurately constructing a causal graph poses a significant challenge. 2) Inverse Propensity Score (IPS) \cite{schnabel2016recommendations, swaminathan2015self}, which adjusts the data distribution by reweighing the observed samples. While IPS can theoretically achieve unbiasedness, its performance is highly sensitive to propensity configuration and prone to high variance. 3) Doubly Robust Learning (DR) \cite{wang2019doubly, guo2021enhanced, dai2022generalized, ding2022addressing}, which enhances IPS by incorporating error imputation for all user-item pairs. DR  enjoys the doubly robust property where unbiasedness is guaranteed if either the imputed values or propensity scores are accurate.

Encouraged by the promising performance and theoretical advantages of DR, this study opts for the DR  approach. However, we highlight a potential limitation of current DR  methods --- they conduct imputation for all user-item pairs, potentially leading to \textit{poisonous imputation}. In DR,  imputation values rely on the imputation model, which is typically trained on a small set of observed data and extrapolated to the entire user-item pairs. Consequently, it is inevitable for the imputation model to produce inaccurate estimations on certain user-item pairs. Poisonous imputation arises when the imputed values significantly diverge from the truth to such an extent that they negatively impact the debiasing process and could even compromise the model's performance. Upon examining existing DR  methods on real-world datasets, we found that the ratio of poisonous imputation is notably high, often exceeding 35\%. Addressing poisonous imputation is thus essential for the effectiveness of a DR  method.

% In DR-JL, the imputation values relies on the imputation model, while it is usually learned on a small observed data and extrapolate the large whole user-item pairs. It is inevitable that the imputation model would make incurate estimation on some user-item pairs. 
% Poisonous imputation occurs in some user-item pairs where the imputation values deviates from the truth to such an large extent that the imputation makes no progress and even hurt the model performance. In fact, by examining existing DR-JL methods on real-world datasets, we find the ratio of poisonous imputation are quite large --- usually over 40\%. Addressing poisonous imputation is essential for a DR-JL method. 

% Recent years have witnessed a plenty of methods on adderessing selection bias including Invariant Learning, Inverse Propensity Score (IPS) and Doubly Robust Learning (DR). Amaong these methods, DR-JL has attracted great attention due to its encouraging performance and great theoretical properities. Specifically, beyond IPS that ajust training data distribution by reweighing the observed data instances, DR-JL introduces the error imputation for all user-item pairs and enjoys unbiasedness if either the imputation or propensity are accurate. 

% . The imputation model is usually learned on a small portion of observed data while attempts to extrapolate large-scale all user-item pairs. It is highly challenge and even impossible for the imputation model to make accurate prediction on all user-item pairs. 

A straightforward solution to this issue could be to directly identify and eliminate poisonous imputations. However, this is practically infeasible due to the unavailability of ground-truth labels of user preference for the majority of user-item pairs. To address this challenge, we propose a \textbf{C}onservative \textbf{D}oubly \textbf{R}obust strategy (CDR) that constructs a surrogate filtering protocol by scrutinizing the mean and variance of the imputation value. Theoretical analyses demonstrate our CDR achieves lower variance and better tail bound compared to conventional DR. Remarkably, our solution is model-agnostic and can be easily plug-in existing DR  methods. In our experiments, we implemented CDR in four different methods, demonstrating that CDR yields superior recommendation performance and a reduced ratio of poisonous imputation. 

To summarize, this work makes the following contributions:

\begin{itemize}
    \item Exposing the issue of poisonous imputation within existing Doubly Robust methods in Recommendation Systems.
    \item Proposing a Conservative Doubly Robust strategy (CDR) that mitigates the problem of poisonous imputation through examination of the mean and variance of the imputation value.
    \item Performing rigorous theoretical analyses and conducting extensive empirical experiments to validate the effectiveness of CDR.
\end{itemize}

\section{Analyses over Doubly Robust Learning}
In this section, we first formulate the task of recommendation debiasing (Sec. 2.1), and then present some background of doubly robust learning (Sec. 2.2). Finally, we identify the issue of poisonous imputation on existing DR  methods (Sec. 2.3). 

\subsection{Task Formulation}
Suppose we have a recommender system composed of a user set $\mathcal U$ and an item set $\mathcal I$. Let $\mathcal{D = U \times I}$ denote the set of all user-item pairs. Further, let $r_{ui}\in \mathbb R$ be the ground-truth label (\eg rating) for a user-item pair $(u,i)$, indicating how the user likes the item; and $\hat r_{ui}$ be the corresponding predicted label from a recommendation model. The collected historical rating data can be notated as a set $\mathbf{R}^o=\{r_{ui}|o_{ui}=1\}$, where $o_{ui}$ denotes whether the rating of a user-item pair $(u,i)$ is observed. The goal of a RS is to accurately predict user preference and accordingly identify items that align with users' tastes. The ideal loss for training a recommendation model can be formulated as follow:

\begin{equation}
    \mathcal{L}_{ideal} \ =\ \mid\mathcal{D}\mid^{-1} \sum_{(u,i) \in \mathcal{D}} e_{ui}
\end{equation}
Where $e_{ui}$ denotes the prediction error between $r_{ui}$ and $\hat r_{ui}$, \eg $e_{ui}=|r_{ui}-\hat r_{ui}|^2$ with RMSE loss or $e_{ui}=-r_{ui}\log(\hat r_{ui})-(1-r_{ui})\log(1-\hat r_{ui})$ with BCE loss. However, only a small portion of $r_{ui}$ is observed in RS, rendering the ideal loss non-computable. Moreover, the challenge is further accentuated by the presence of selection bias, as the observed data might not faithfully represent the entirety of user-item pairs. For instance, samples with higher ratings are more likely to be observed \cite{marlin2009collaborative}. Utilizing a naive estimator that calculates directly on the observed data with $\mathcal{L}_{naive}=\mid\mathcal{D}\mid^{-1} \sum_{(u,i) \in \mathcal{D}} o_{ui}e_{ui}$ would yield a biased estimation \cite{schnabel2016recommendations}. Hence, the exploration for a suitable surrogate loss towards unbiased estimation of the ideal loss is ongoing. 

\subsection{Existing Estimators}
Now we review two typical estimators for addressing selection bias.

\textit{Inverse Propensity Score Estimator (IPS) \cite{schnabel2016recommendations}.} The IPS estimator aims to adjust the training distribution by reweighing the observed instances as:
\begin{equation}
    \mathcal{L}_{IPS} \ =\ \mid\mathcal{D}\mid^{-1} \sum_{(u,i) \in \mathcal{D}} \frac{o_{ui}e_{ui}}{\hat{p}_{ui}} 
\end{equation}
where $\hat{p}_{ui}$ is an estimation of the propensity score $p_{ui} = \mathbb P(o_{ui}=1)$. The bias and variance of IPS estimator can be written as:

\begin{equation}
    \begin{split}
    Bias[\mathcal{L}_{IPS}] &=\mathbb |E_o[\mathcal{L}_{IPS}]-\mathcal{L}_{Ideal}| \\ &= \mid\mathcal{D}\mid^{-1} \lvert \sum_{(ui) \in \mathcal{D}} \frac{(p_{ui} - \hat{p}_{ui})}{\hat{p}_{ui}}{e_{ui}} \rvert \\
    Var[\mathcal{L}_{IPS}] & = \mathbb E_o[(\mathcal{L}_{IPS}-\mathbb E_o[\mathcal{L}_{IPS}])^2] \\ &=\mid\mathcal{D}\mid^{-2} \sum_{(u,i) \in \mathcal{D}} \frac{p_{ui}(1 - p_{ui})}{\hat{p}_{ui}^2}{{e}_{ui}^2}
    \end{split}
\end{equation}
Once the $\hat{p}_{ui}$ reaches its ideal value (\ie $\hat{p}_{ui}=p_{ui}$), the IPS estimator could provide an unbiased estimation of the ideal loss (\ie $\mathbb E_o[\mathcal{L}_{IPS}]=\mathcal{L}_{Ideal}$). 

\textit{Doubly Robust Estimator (DR) \cite{wang2019doubly}.} DR  augments IPS by introducing the error imputation with the following loss:
\begin{equation}
    \mathcal{L}_{DR} \ =\ \mid\mathcal{D}\mid^{-1} \sum_{(u,i) \in \mathcal{D}} (\hat{e}_{ui} + \frac{o_{ui}(e_{ui} - \hat{e}_{ui})}{\hat{p}_{ui}})
\end{equation}
where $\hat e_{ui}$ represents the imputed error, derived from a specific imputation model that strives to fit the predicted error. Recent work \cite{guo2021enhanced} has established the bias and variance of DR  as follow:

\begin{equation}
    \begin{split}
        \label{eq:drb}
    Bias[\mathcal{L}_{DR}] & = \mid\mathcal{D}\mid^{-1} \lvert \sum_{(u,i) \in \mathcal{D}} \frac{(p_{ui} - \hat{p}_{ui})}{\hat{p}_{ui}}{(e_{ui} - \hat{e}_{ui})} \rvert \\
    Var[\mathcal{L}_{DR}] & = \mid\mathcal{D}\mid^{-2} \sum_{(u,i) \in \mathcal{D}} \frac{p_{ui}(1 - p_{ui})}{\hat{p}_{ui}^2}{(\hat{e}_{ui} - e_{ui})^2}
    \end{split}
\end{equation}
As can be seen, DR  change the bias term for each $(u,i)$ from $ \frac{(p_{ui} - \hat{p}_{ui})}{\hat{p}_{ui}}{e_{ui}}$ in IPS to $\frac{(p_{ui} - \hat{p}_{ui})}{\hat{p}_{ui}}{(e_{ui} - \hat{e}_{ui})}$ and the variance from $\frac{p_{ui}(1 - p_{ui})}{\hat{p}_{ui}^2}{{e}_{ui}^2}$ to $\frac{p_{ui}(1 - p_{ui})}{\hat{p}_{ui}^2}{(\hat{e}_{ui} - e_{ui})^2}$. DR  enjoys the doubly robust property that if either $\hat{p}_{ui}=p_{ui}$ or $e_{ui} = \hat{e}_{ui}$ holds, $\mathcal{L}_{DR}$ could be an unbiased estimator (\ie $Bias[\mathcal{L}_{DR}]=0$). This advantageous property typically results in DR  being less biased than IPS in practice, empirically leading to superior performance.

\subsection{Limitation of DR }

\begin{table}[]
    \centering
    \begin{tabular}{cccc}
         \toprule
         & Coat & Yahoo & KuaiRand  \\ \hline
         DR-JL & 45.9\% & 41.9\% & 38.8\% \\ \hline
         MRDR & 48.1\% & 43.1\% & 41.2\% \\\hline
         DR-BIAS & 44.1\% & 40.4\% & 39.2\% \\\hline
         TDR & 42.3\% & 36.2\% & 36.3\% \\ \bottomrule
    \end{tabular}
    \caption{The proportion of "poisonous imputation" in three different datasets using four typical DR methods.}
    \label{tab_poisonous}
\end{table}

From the eq.(\ref{eq:drb}), we can conclude that the accuracy of imputation $\hat e_{ui}$ is of highly importance --- both the bias and variance term are correlated with $|\hat e_{ui}-e_{ui}|$. Indeed, if the imputed error $\hat e_{ui}$ diverges significantly from the predicted error $e_{ui}$ such that $|\hat e_{ui}-e_{ui}|>e_{ui}$, the imputation $\hat e_{ui}$ becomes counterproductive. Particularly, imputing $\hat e_{ui}$ for the user-item pair $(u,i)$ results in increased bias and variance, rather than reduced. We denote this phenomenon as poisonous imputation:
\begin{definition}[Poisonous Imputation]
    For any user-item pair $(u,i)$, the imputation $\hat e_{ui}$ is considered as a poisonous imputation if $|\hat e_{ui}-e_{ui}|>e_{ui}$. 
    \end{definition}

In practical RS, given that the imputation model is typically trained on a limited set of observed data and generalized to the entire user-item pairs, poisonous imputation is frequently encountered. To provide empirical evidence for this point, we conducted an empirical analysis on four representative DR methods (DR-JL \cite{wang2019doubly}, MRDR \cite{guo2021enhanced}, DR-BIAS \cite{dai2022generalized}, and TDR \cite{li2023tdr}) across three real-world debiasing datasets (YahooR3, Coat, and KuaiRand). These DR  methods were finely trained on the biased training data, after which $e_{ui}$ and $\hat e_{ui}$ were calculated for the user-item pairs in the test data where ground-truth ratings are accessible. The proportion of poisonous imputation is reported in Table \ref{tab_poisonous}. \textbf{Surprisingly, the ratio of poisonous imputation is considerably high, often exceeding 35\% across all datasets and baseline models.} It is noteworthy that even though DR  generally exhibits superior performance over IPS, a substantial amount of poisonous imputation still exists. The issue of poisonous imputation is particularly severe, thereby warranting attention and resolution.

\section{METHODOLOGY}
In this section, we first introduce the proposed conservative doubly robust strategy, and then conduct theoretical analyses to validate its merits. 

\begin{figure}
  \includegraphics[width=0.98\linewidth]{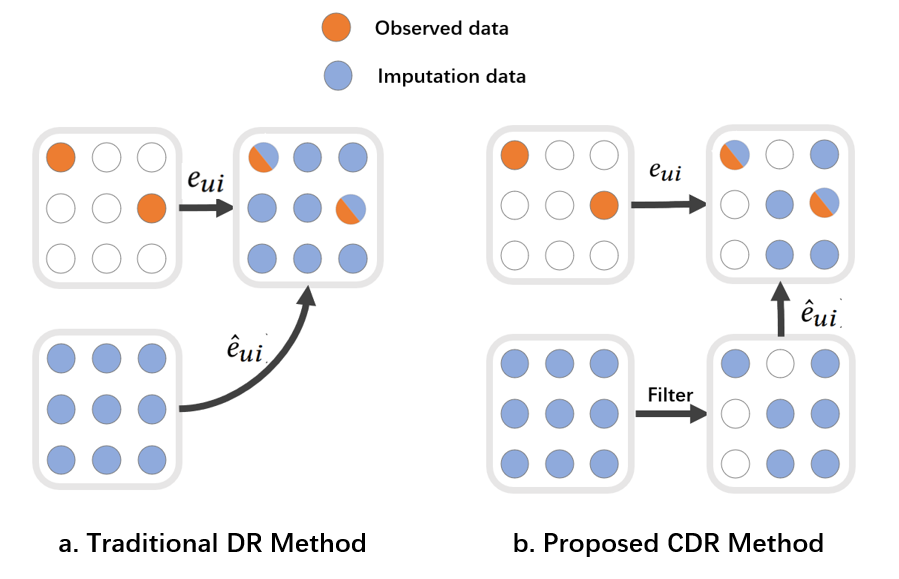}
  \caption{Illustration of how our CDR improves the traditional DR methods --- leveraging a filter protocol to remove the poisonous imputation that may hurt debiasing.}
  \label{fig_CDR_method}
\end{figure}

\subsection{Conservative Doubly Robust Learning}

Considering the widespread occurrence of poisonous imputation, we contend that performing imputation blindly on all user-item pairs, as is customary with current methods, may not be the optimal strategy. Instead, it would be more effective to adopt a conservative and adaptive imputation approach that focuses on user-item pairs which confer benefits while excluding those leading to poisonous imputation. As previously discussed, the ideal filtering protocol involves comparing $|\hat e_{ui}-e_{ui}|$ with $e_{ui}$. If $|\hat e_{ui}-e_{ui}|<e_{ui}$, the imputation should be retained as it could potentially reduce both variance and bias; if not, it implies a poisonous imputation which should be discarded. However, this approach is impractical as the ground-truth labels are typically inaccessible in real-world scenarios and $e_{ui}$ cannot be calculated. As such, an alternative filtering protocol is necessary.

Towards this end, we propose a Conservative Doubly Robust (CDR) strategy in this work that filters imputation by examining the mean and variance of $\hat e_{ui}$. The foundation of CDR is based on the following important lemma:

\newtheorem{lem}{Lemma}
\begin{lem}
	\label{la1}

    Given that $\hat e_{ui}$ and $e_{ui}$ are independently drawn from two Gaussian distributions $ \mathcal N(\hat \mu_{ui},\hat \sigma^2_{ui})$ and $\mathcal N(\mu_{ui},\sigma^2_{ui})$, where $\hat \mu_{ui}$, $\mu_{ui}$, $\hat \sigma_{ui}$, $\sigma_{ui}$ are bounded with $|\hat \mu_{ui}-\mu_{ui}|\leq \varepsilon_\mu$, $|\hat \sigma_{ui}^2-\sigma_{ui}^2|\leq \varepsilon_\sigma^2$, $2\varepsilon_\mu \leq \hat \mu_{ui}$, $m_\mu\leq \hat \mu_{ui} \leq M_\mu$ and $m_\sigma\leq \hat \sigma_{ui} \leq M_\sigma$, for any confidence level $\rho$ ($0\leq \rho \leq 1$), the condition $\mathbb P(|\hat e_{ui}-e_{ui}|<e_{ui})\geq \rho$ holds if 
    \begin{equation}
        \label{eq:pro}
        % \frac{\hat \sigma_{ui}}{\hat \mu_{ui}}<\bigg (\sqrt{5}\Phi^{-1}(\rho)+\frac{2\hat \mu_{ui}\varepsilon_\sigma}{\hat \sigma_{ui}(\sqrt{5}\sigma_{ui}+2\varepsilon_\sigma)}+\frac{2\sqrt{5}\varepsilon_\mu}{\sqrt{5}\sigma_{ui}+2\varepsilon_\sigma}\bigg )^{-1}
        \frac{{{{\hat \sigma }_{ui}}}}{{{{\hat \mu }_{ui}}}} <  {\bigg (\sqrt 5 {\Phi ^{ - 1}}(\rho ) + \frac{{2{M_\mu }{\varepsilon _\sigma }}}{{{m_\sigma }(\sqrt 5 {m_\sigma } + 2{\varepsilon _\sigma })}} + \frac{{2\sqrt 5 {\varepsilon _\mu }}}{{\sqrt 5 {m_\sigma } + 2{\varepsilon _\sigma }}}\bigg )^{ - 1}}
    \end{equation}
    where $\Phi^{-1}(.)$ denotes the inverse of CDR of the standard normal distribution.
	% Given that $\hat e_{ui}$ and $e_{ui}$ are independently DR-JLawn from two Gaussian distributions $ \mathcal N(\mu_{1,ui},\sigma_{1,ui})$ and $\mathcal N(\mu_{2,ui},\sigma_{2,ui})$, let $\mu_{1,ui}$, $\mu_{2,ui}$, $\sigma_{1,ui}$, $\sigma_{2,ui}$ are bounded with $|\mu_{1,ui}-\mu_{2,ui}|\leq \epsilon_\mu$, $|\sigma_{1,ui}^2-\sigma_{2,ui}^2|\leq \epsilon_\sigma^2$, for any confidence $\rho$ ($0\leq \rho \leq 1$), the condition $\mathbb P(|\hat e_{ui}-e_{ui}|<e_{ui})>\rho$ holds if 
    % \begin{equation}
    %     \frac{\sigma_{1,ui}}{\mu_{1,ui}}<(\sqrt(5)\Phi^{-1}(\rho)+\frac{2\mu_{1,ui}\epsilon_\sigma}{\sqrt(5)}\sigma_{1,ui})
    % \end{equation}
\end{lem}

The proof of the lemma is included in the appendix A. This lemma indicates that through the formulation of a distribution hypothesis for $\hat e_{ui}$ and $e_{ui}$, the evaluation of poisonous imputation can be reframed as a scrutiny of the mean and variance of $\hat e_{ui}$. The hypothesis presented in the lemma is practical. On one hand, we hypothesize that the distribution of $\hat e_{ui}$ approximates that of $e_{ui}$ (i.e., $|\hat \mu_{ui}-\mu_{ui}|\leq \varepsilon_\mu$, $|\hat \sigma_{ui}^2-\sigma_{ui}^2|\leq \varepsilon_\sigma^2$, $2\varepsilon_\mu \leq \hat \mu_{ui}$), a supposition that naturally follows since the imputation model endeavors to fit $e_{ui}$. On the other hand, we opt to employ the Gaussian distribution for analysis. This choice is informed by its widespread usage in statistical inference, as well as its standing as a second-order Taylor approximation of any distribution. While more complex distributions might yield more precise results, \eg considering higher-order moments, the analytical complexity and computational burden would significantly increase. Our empirical findings indicate that the Gaussian distribution suffices to deliver superior performance.

In fact, our proposed filtering protocol (inequality (\ref{eq:pro})) is intuitively appealing due to three observations: 1) A larger value of $\hat \sigma_{ui}$ makes the preservation of the imputation less likely. This is consistent with the understanding that a higher variance implies a less reliable prediction, thus making it more susceptible to discarding. 2) A larger value of $\hat \mu_{ui}$ makes the preservation of the imputation more likely. This can be rationalized by the notion that if the error $e_{ui}$ is large, the imputation is safer as it is more difficult to exceed $2e_{ui}$. 3) Larger values of $\varepsilon_{\mu}$ and $\varepsilon_{\sigma}$ increase the likelihood of filtering the imputation. Larger values for these parameters suggest a more significant distributional gap between $e_{ui}$ and $\hat e_{ui}$, thereby necessitating more conservative filtering.

\textbf{Instantiation of CDR.} CDR can be incorporated into various DR  methods by leveraging an additional filtering protocol. This protocol consists of two steps: 

1) Estimation of $\hat \mu_{ui}, \hat \sigma_{ui}$: We utilize the Monte Carlo Dropout method \cite{gal2016dropout} for estimating the mean and variance of the imputation, owing to its generalization and easy implementation.  Specifically, we apply dropout 10 times on the imputation model (\ie randomly omitting 50\% of the dimensions of embeddings) and then calculate the mean and variance of $\hat e_{ui}$ from the dropout model. To ensure a fair comparison, we should note that dropout is only employed during the calculation of $\hat \mu_{ui}, \hat \sigma_{ui}$, and not during the training of the imputation model. 

2) Filtering based on the condition $\frac{\hat \sigma_{ui}}{\hat \mu_{ui}}<\eta$: Note that the right-hand side of inequality (\ref{eq:pro}) involves complex computation and five parameters. To simplify our implementation, we re-parameterize the right-hand side of the inequation as a hyperparameter $\eta$. This parameter $\eta$ can be interpreted as an adjusted threshold that directly modulates the strictness of the filtering process.

With the above filtering protocol, the CDR estimator can be formulated as:
\begin{equation}
      \mathcal{L}_{CDR}  =|{\mathcal D}{|^{ - 1}}\sum\limits_{(u,i) \in {\mathcal D}} {(\frac{{{o_{ui}}{e_{ui}}}}{{{{\hat p}_{ui}}}} + {\gamma _{ui}}{{\hat e}_{ui}}(1 - \frac{o_{ui}}{{{{\hat p}_{ui}}}}))} 
\end{equation}
where $\gamma_{ui}\in \{0,1\}$ indicate whether the imputation $\hat e_{ui}$ is retained.

\subsection{Theoretical Analyses}
In order to elucidate the advantages of the Conservative Doubly Robust (CDR) strategy, we present the following lemma:
\begin{lem}
	\label{la2}
    Given the imputed errors $\hat e_{ui}$, estimated propensity scores $\hat p_{ui}$, and the retention of the imputation $\gamma_{ui}$, the bias and variance of the CDR estimator can be expressed as follows:
\begin{equation}
    \begin{split}
Bias[ \mathcal{L}_{CDR}]&=\frac{1}{|{\mathcal D}|} \lvert \! \! \! \sum\limits_{(u,i) \in {\mathcal D}}\! \! \! {\frac{{({p_{ui}}\! - \!{{\hat p}_{ui}})}}{{{{\hat p}_{ui}}}}({\gamma _{ui}}} ({e_{ui}} \! -\!  {{\hat e}_{ui}}) \! + \! (1\!  -\!  {\gamma _{ui}}){e_{ui}})\rvert \\
Var[ \mathcal{L}_{CDR}]&=\frac{1}{|{\mathcal D}|^2}\! \! \! \sum\limits_{(u,i) \in {\mathcal D}} \! \! \! {\frac{{{p_{ui}}(1 \!-\! {p_{ui}})}}{{\hat p_{ui}^2}}} ({\gamma _{ui}}{({{\hat e}_{ui}}\! -\! {e_{ui}})^2}\! + \!(1\! - \!{\gamma _{ui}})e_{ui}^2)
    \end{split} 
\end{equation}
With probability $1-\kappa$, the deviation of the CDR estimator from its expectation has the following tail bound: 
\begin{equation}
    |\mathcal{L}_{CDR}\!-\!\mathbb E_o[\mathcal{L}_{CDR}]|\! \leq \! \sqrt {\frac{{\log \left( {\frac{2}{\kappa }} \right)}}{{2|{\mathcal D}{|^2}}}\! \! \!\sum\limits_{u,i \in {\mathcal D}} \! \! \!{({\gamma _{ui}}\frac{{{{({e_{ui}}  \! -  \! {{\hat e}_{ui}})}^2}}}{{\hat p_{ui}^2}} \! +  \! (1  \! -  \! {\gamma _{ui}})\frac{{e_{ui}^2}}{{\hat p_{ui}^2}})} } 
\end{equation}
\end{lem}
The proof is presented in appendix B. CDR can be understood as an integration of IPS and DR . If 
$|\hat e_{ui}-e_{ui}|>e_{ui}$, CDR will filter out the poisonous imputation and regress to IPS, as IPS demonstrates superior bias and variance properties compared to DR . Otherwise, CDR will retain the imputation, benefiting from the merits of DR . Indeed, CDR has the following advantages:

\begin{corollary}
\label{co}
Under the condition of Lemma \ref{la1} and ${\varepsilon _\mu } \ll {\hat \mu _{ui}}, \varepsilon _\mu ^2 \ll \hat \sigma _{ui}^2$, with a proper filtering threshold $\eta$, CDR enjoys better variance and tail bound than IPS and DR. 
\end{corollary}
The proof is presented in appendix C. This corollary substantiates the superiority of CDR, thereby yielding better recommendation performance. We will empirical validate it in the following section.

\begin{table*}[t]
    \caption {Performance comparison between our CDR with other baselines on three real-world datasets. The best result in that column is bolded and the runner-up is underlined. We incorporate CDR into four baseline models and report the relative improvements gained by employing CDR compared to the respective baseline. }
    \label{tab_main_result}
    \begin{tabular}{l|ccccccccc}
        \toprule
        \multirow{2}{*}{Method} & \multicolumn{3}{c}{Coat} & \multicolumn{3}{c}{Yahoo} & \multicolumn{3}{c}{KuaiRand} \\
        % \cline{2-10}
        &AUC &NDCG@5  &Recall@5   &AUC &NDCG@5 &Recall@5   &AUC &NDCG@5  &Recall@5 \\
        \hline
        MF
        &0.7053 &0.6025 &0.6173 
        &0.6720	&0.6252	&0.7155
        &0.5432	&0.2932	&0.2905
        \\
        IPS
        &0.7144 &0.6173 &0.6267
        &0.6785	&0.6345	&0.7214
        &0.5446	&0.2987	&0.2987
        \\
        CVIB
        &0.7230 &0.6278 &0.6347
        &0.6811 &0.6482 &0.7229
        &0.5512 &0.3099 &0.3027
        \\
        INV 
        &0.7416 &0.6394 &0.6542 
        &0.6767	&0.6443	&0.7251
        &0.5465	&0.3081	&0.3013
        \\
        TDR 
        &0.7388 &0.6378 &0.6525
        &0.6789	&0.6436	&0.7269
        &0.5523	&0.3088	&0.3026
        \\ \hline
        EIB
        &0.7225	&0.6288	&0.6382 
        &0.6844	&0.6427	&0.7241
        &0.5456	&0.3010	&0.2938
        \\
        \cellcolor{lightblue}EIB+CDR
        &\cellcolor{lightblue}\underline{0.7509}	&\cellcolor{lightblue}0.6533	&\cellcolor{lightblue}0.6608
        &\cellcolor{lightblue}\underline{0.6909}	&\cellcolor{lightblue}0.6549	&\cellcolor{lightblue}0.7310
        &\cellcolor{lightblue}0.5510	&\cellcolor{lightblue}0.3087	&\cellcolor{lightblue}0.2975
        \\
        impv\% 
        &+3.93\% &+3.90\% &+3.54\%
        &+0.95\% &+1.90\% &+0.95\%
        &+0.99\% &+2.56\% &+1.26\%
        \\ \hline
        DR-JL
        &0.7286	&0.6271	&0.6355
        &0.6834	&0.6474	&0.7236
        &0.5485	&0.2967	&0.2924
        \\
        \cellcolor{lightblue}DR+CDR
        &\cellcolor{lightblue}0.7502	&\cellcolor{lightblue}\underline{0.6557}	&\cellcolor{lightblue}\underline{0.6658} 
        &\cellcolor{lightblue}0.6881	&\cellcolor{lightblue}0.6558	&\cellcolor{lightblue}0.7307
        &\cellcolor{lightblue}\underline{0.5540}	&\cellcolor{lightblue}\underline{0.3153}	&\cellcolor{lightblue}\underline{0.3045}
        \\
        impv\% 
        &+2.96\% &+4.56\% &+4.77\%
        &+0.69\% &+1.31\% &+0.98\%
        &+1.00\% &+6.27\% &+4.14\%
        \\ \hline
        MRDR
        &0.7319	&0.6317	&0.6447
        &0.6829	&0.6484	&0.7243
        &0.5503	&0.3041	&0.2949
        \\
        \cellcolor{lightblue}MRDR+CDR
        &\cellcolor{lightblue}0.7508	&\cellcolor{lightblue}0.6520	&\cellcolor{lightblue}0.6587
        &\cellcolor{lightblue}0.6879	&\cellcolor{lightblue}\textbf{0.6571}	&\cellcolor{lightblue}\underline{0.7311}
        &\cellcolor{lightblue}\textbf{0.5547}	&\cellcolor{lightblue}\textbf{0.3167}	&\cellcolor{lightblue}\textbf{0.3078}
        \\
        impv\% 
        &+2.58\% &+3.21\% &+2.17\%
        &+0.73\% &+1.34\% &+0.94\%
        &+0.80\% &+4.14\% &+4.48\%
        \\ \hline
        DR-BIAS
        &0.7424	&0.6408	&0.6578
        &0.6860	&0.6486	&0.7269
        &0.5478	&0.3024	&0.2952
        \\
        \cellcolor{lightblue}DR-BIAS+CDR
        &\cellcolor{lightblue}\textbf{0.7513}	&\cellcolor{lightblue}\textbf{0.6567}	&\cellcolor{lightblue}\textbf{0.6678}
        &\cellcolor{lightblue}\textbf{0.6912}	&\cellcolor{lightblue}\underline{0.6565}	&\cellcolor{lightblue}\textbf{0.7323}
        &\cellcolor{lightblue}0.5533	&\cellcolor{lightblue}0.3098	
        &\cellcolor{lightblue}0.3048
        \\
        impv\% 
        &+1.20\% &+2.48\% &+1.52\%
        &+0.76\% &+1.22\% &+0.74\%
        &+1.00\% &+2.45\% &+3.25\%
        \\
        \bottomrule
    \end{tabular}
\end{table*}

\section{Experiments}

In this section, we designed experiments to test the performance of the proposed method on three real-world datasets. Our aim was to answer the following four research questions:

\begin{itemize}
	\item [\textbf{RQ1:}] Does the proposed CDR improve the debiasing performance? 
	\item [\textbf{RQ2:}] Does CDR indeed reduce the ratio of poisonous imputation in DR? 
	\item [\textbf{RQ3:}] How does the hyperparameter $\eta$ (filtering threshold) affect debiasing performance?
    \item [\textbf{RQ4:}] Does CDR incur much more computational time?
\end{itemize}

% RQ1: Does the proposed CDR improve the debiased recommendation performance?

% RQ2: Does MCDR really reduce the probability of misjudging imputation model imputed values?

% RQ3: How does the performance of the model vary with different threshold values?

% RQ4: Does MCDR significantly increase execution time?

\subsection{Experimental Setup}
\textbf{Datasets.}
To evaluate the performance of debiasing methods on real-world datasets, The ground-truth unbiased data are necessary. We closely refer to previous studies\cite{schnabel2016recommendations, wang2019doubly, guo2021enhanced, gao2022kuairand}, and use the following three benchmark datasets: \textbf{Coat}, \textbf{Yahoo!R3} and \textbf{KuaiRand-Pure}.
All three datasets consist of a biased dataset, collected from normal user interactions, and an unbiased dataset collected from random logging strategy. Specifically, \textbf{Coat} includes 6,960 biased ratings and 4,640 unbiased ratings from 290 users for 300 items; 
\textbf{Yahoo!R3} comprises 54,000 unbiased ratings and 311,704 biased ratings from 15,400 users for 1,000 items; while
\textbf{KuaiRand} includes 7,583 videos and 27,285 users, containing 1,436,609 biased data and 1,186,059 unbiased data. Following recent work \cite{chen2021autodebias},
we regard the biased data as training set, and utilize the unbiased data for model validation (10\%) and evaluation (90\%). Also, the ratings are binarized with threshold 3. That is, the observed rating value larger than 3 is labeled as positive, otherwise negative.

%On average, each user interacts with more than 1,000 videos over a period of four weeks.

\textbf{Baselines.}
We validate the effectiveness of CDR on four baselines including three benchmark DR methods and one classical baseline just based on imputation:
\begin{itemize}
    \item \textbf{EIB\cite{steck2013evaluation}:} the classical baseline that relies on data imputation for tackling selection bias.
    \item \textbf{DR-JL\cite{wang2019doubly}:} the basic doubly robust learning strategy that employs both propensity and imputation for recommendation debiasing. In DR-JL, the imputation is learned by minimizing the error deviation on observed data.
    \item \textbf{MRDR\cite{guo2021enhanced}:} the method improves DR-JL by considering the variance reduction for learning imputation model.
    \item \textbf{DR-BIAS\cite{dai2022generalized}:} the novel strategy that learns imputation with balancing the variance and bias.
\end{itemize}
We also compare the methods with:
\begin{itemize}
    \item \textbf{Base Model:} the basic recommendation model without employing any debiasing strategy.
    \item \textbf{IPS\cite{schnabel2016recommendations}:} the strategy that addresses bias via weighing the observed data with the inverse of the propensity.
    \item \textbf{INV\cite{wang2022invariant}:} the state-of-the-art debiasing method that leverages causal graph to disentangle the invariant preference and variant factors from the observed data.
    \item \textbf{TDR\cite{li2023tdr}:} the state-of-the-art DR method that learns imputation with a parameterized imputation model and a non-parameter strategy. Here we do not implement CDR in TDR due to its high complexity. Nevertheless, our experiments show that even CDR is plug-in the basic DR-JL, it could outperform TDR.
\end{itemize}
Also, for fair comparison, we closely refer to recent work \cite{chen2021autodebias} and take the widely used Matrix Factorization (MF) \cite{koren2009matrix} as the base recommendation model.

% the proposed methods with the following baselines: Base Model \cite{koren_matrix_2009}, IPS \cite{schnabel_recommendations_2016}, DR-JL \cite{wang_doubly_2019}, EIB \cite{marlin_collaborative_2012, steck_evaluation_2013}, MRDR \cite{guo_enhanced_2021}, INV \cite{wang_invariant_2022}, TDR \cite{li_tdr-cl_2023} and  dr-bias \cite{dai_generalized_2022}. Furthermore, we leverage Naive Bayes with Laplace smoothing to obtain the propensity.

% We take the widely used Matrix Factorization (MF) \cite{koren_matrix_2009} as the base model, and

\textbf{Metrics.}
We employed three concurrent metrics, namely, Area Under the Curve (AUC), Recall (Recall@5) and Normalized Discounted Cumulative Gain (NDCG@5) to assess debiasing performance. NDCG@K evaluates the quality of recommendations by taking into account the importance of each item's position, based on discounted gains. 
\begin{equation}
\begin{split}
    DCG_u@K &= \sum_{(u,i)\in \mathcal{D}_{test}} \frac{I(\hat{z}_{u,i} <= K)}{log(\hat{z}_{u,i} + 1)} \\
    NDCG@K &= \frac{1}{\mid \mathcal{U} \mid} \sum_{u \in \mathcal{U}} \frac{DCG_u@K}{IDCG_u@K}
\end{split}
\end{equation}

where IDCG represents the ideal DCG, $\mathcal{D}_{test}$ denotes the test data, $\hat{z}_{u,i}$ represents the position of item $i$ within the recommended rank for user $u$.

Recall@K measures the number of recommended items that are likely to be interacted with by the user within top $K$ items.
\begin{equation}
\begin{split}
    Recall_u@K &= \frac{\sum_{(u,i)\in \mathcal{D}_{test} I(\hat{z}_{u,i} <= k)}}{\mid \mathcal{D}^u_{test} \mid} \\
    Recall@K &= \frac{1}{\mid \mathcal{U} \mid} \sum_{u \in \mathcal{U}} Recall_u@K
\end{split}
\end{equation}
where $\mathcal{D}^u_{test}$ indicates all ratings of the user $u$ in dataset $\mathcal{D}_{test}$.

\textbf{Experimental details.}  Our experiments were conducted on PyTorch, utilizing Adam as the optimizer. We fine-tuned the learning rate within \{0.005, 0.01, 0.05, 0.1\}, weight decay within \{1e - 5, 5e - 5, 1e - 4, 5e - 4, 1e - 3, 5e - 3, 1e - 2\}, threshold's parameter $\eta$ within \{0.1, 0.5, 1, 3, 5, 7, 10, 50\}, and batch size within \{128, 256, 512, 1024, 2048\} for Coat , \{1024, 2048, 4096, 8192, 16384\} for Yahoo!R3 and \{ 2048, 4096, 8192, 16384, 32768\} for KuaiRand. The hyperparameters of all the baselines are finely tuned in our experiments or referred to the orignal paper. The code is available at $https://github.com/CrazyDumpling/CDR\_CIKM2023$.

\begin{table*}
    \caption{Empirical runtime (s) comparison on Coat, Yahoo and KuaiRand datasets.}
    \label{tab_run_time}
    \begin{tabular}{c|cc|cccc|cccc}
        \toprule
        Datasets & MF & IPS & EIB & DR-JL & MRDR & DR-bias & EIB+CDR & DR-JL+CDR & MRDR+CDR & DR-bias+CDR \\ \hline
        Coat & 33.87 & 36.72 & 112.80 & 136.72 & 131.69 & 138.23 & 135.62 & 147.31 & 153.28 & 149.31\\ \hline
        Yahoo & 59.34 & 68.91 & 542.35 & 632.79 & 687.34 & 678.28 & 643.21 & 732.13 & 706.39 & 714.32 \\ \hline
        KuaiRand & 834.13 & 1034.24 & 5018.23 & 6390.25 & 6246.36 & 6421.56 & 7124.54 & 6893.49 & 7154.83 & 7245.71 \\ \bottomrule
    \end{tabular}
\end{table*}

\subsection{Performance Comparison (RQ1)}
Table \ref{tab_main_result} presents performance comparison of our with other Baselines. We draw the following observations: 

1) CDR consistently boosts the recommendation performance on four baselines and three benchmark datasets. Especially in KuaiRand, the improvement is impressive --- achieving average 0.95\%, 3.86\%, 3.28\% improvement in terms of AUC, NDCG and Recall respectively. This result validates that our filtering protocol is effective, which could indeed filter the harmful imputation. We will further validate this point in the next experiment.

2) By comparing CDR with other baselines, we can find the best performance always achieved by CDR. CDR is simple but achieves SOTA performance.    

\subsection{Study on the Poisonous Imputation (RQ2)}
\begin{figure*}
  \label{fig_poisonous}
  \includegraphics[width=\textwidth]{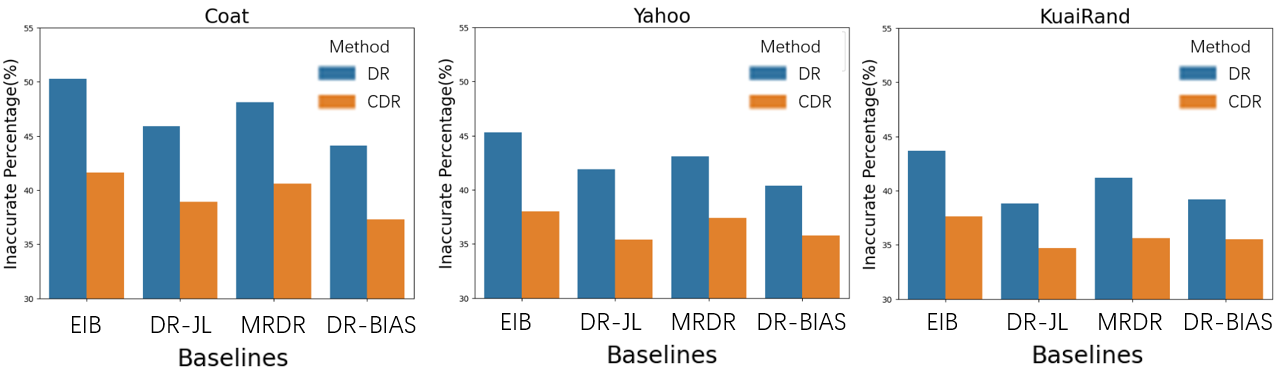}
  \caption{The percentage(\%) of "poisonous imputation" in three different datasets using the original EIB, DR-JL, MRDR and DR-BIAS methods, as well as the improved methods with integrating CDR.}
  \label{fig_inaccurate_per}
\end{figure*}

To further validate the effectiveness of CDR, we conducted empirical study on the ratio of the poisonous imputation. We finely trained compared methods on the biased training data, and then compared $|e_{ui}-\hat e_{ui}|$ with $e_{ui}$ for the user-item pairs in the test data where ground-truth ratings are accessible. The results are presented in Figure \ref{fig_inaccurate_per}.

As can be seen, CDR consistently has lower ratio of poisonous imputation than its corresponding baselines over three datasets. This result clearly validate that the proposed filter is reasonable and can remove a certain ratio of poisonous imputation. As such, CDR achieves better debiasing performance than DR. 

\subsection{ Effect of Hyperparameter $\eta$ (RQ3)}
\begin{figure*}
  \includegraphics[width=\textwidth]{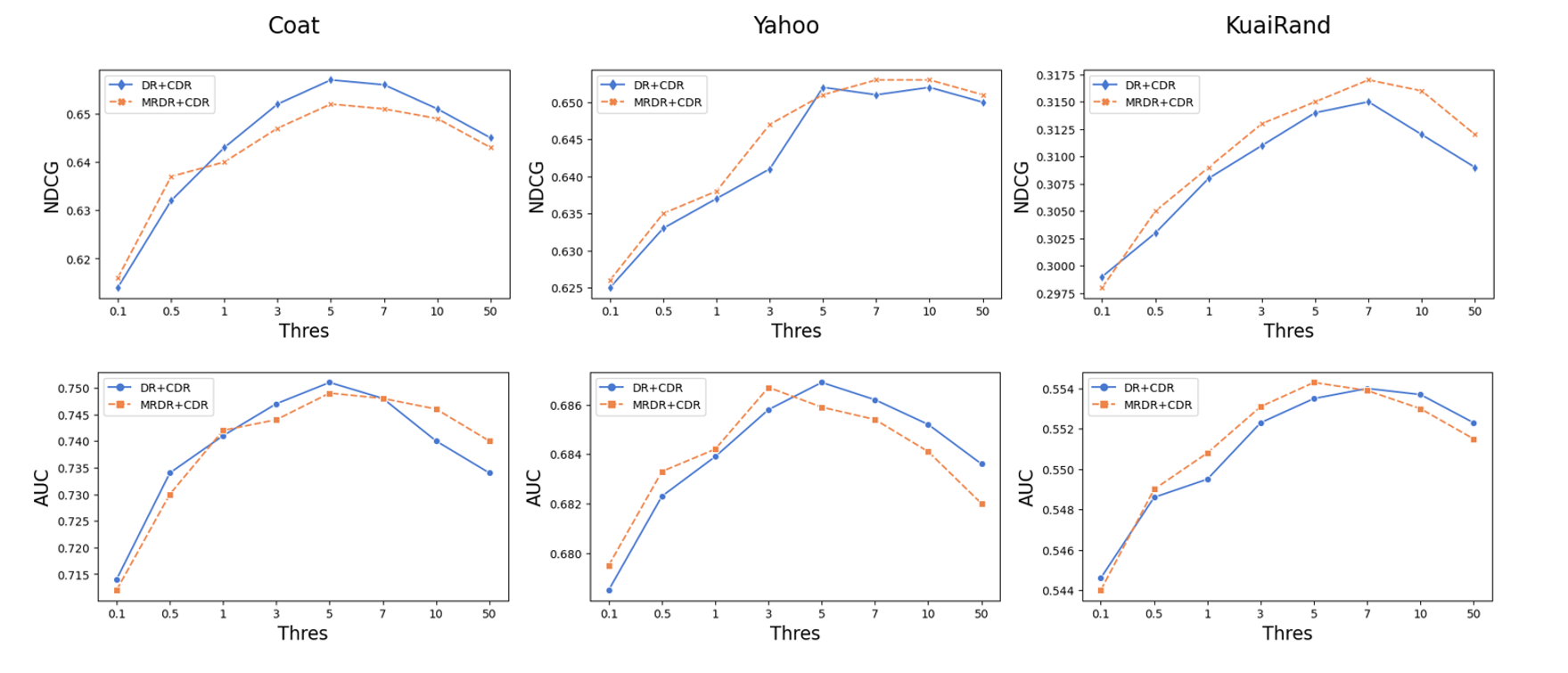}
  \caption{Recommendation performance of CDR with varying threshold $\eta$ on three datasets.}
  \label{fig_thres}
\end{figure*}

The hyperparameter $\eta$ serves as an adjusted threshold that directly modulates the strictness of the filtering process. Thus, exploring model performance \wrt $\eta$ could help us to better understand the nature of CDR. In theoretical terms, when $\eta$ approaches 0, this method is equivalent to IPS; when $\eta$ approaches infinity, this method is equivalent to the DR approach. The performance with varying $\eta$ is presented in Figure \ref{fig_thres}.

As can be seen, with $\eta$ increasing, the performance will become better first. The reason is that the larger $\eta$ would bring more imputation. As the threshold $\eta$ is relatively low, the injected imputation is usually confidence, yielding performance improvement. However, when $\eta$ surpasses a certain value, the performance becomes worse with further increase of $\eta$. This can be interpreted by the more inaccurate imputation is injected. poisonous imputation occurs which would deteriorate model performance. Consequently, there exists a trade-off on the selection of $\eta$. Only when $\eta$ is set to a proper value, the model achieves the optimal performance. 

\subsection{ Running Time Comparison (RQ4)}

Additionally, we conducted experiments on the efficiency of CDR compared with other baselines on three datasets: Coat, Yahoo, and KuaiRand. As shown in the table \ref{tab_run_time}, despite CDR introduces multiply times dropout for evaluating the mean and variance of the imputation, it does not incur much more computation burden. The reason can be attributed to the two factors: 1) The calculation of the mean and variance only involves forward propagation, without requiring the time-consuming backward propagation; 2) CDR would filter a certain ratio of the imputation, which make the samples in training reduced, leading to acceleration when training the recommendation model.

\section{Related Work}
In this section, we review the most related work from the following
two perspectives.

\textbf{Debiasing in Recommendation.}
Bias is a critical issue in recommendation systems as it not only hurt recommendation accuracy, but can limit the diversity of recommended items and reinforce unfairness \cite{chen2023bias, wu2022opportunity,gao2023alleviating}. There are various sources of bias found in RS data, such as selection bias \cite{marlin2009collaborative, marlin2012collaborative,chen2018social}, exposure bias \cite{liu2020general,chen2019samwalker,chen2020fast}, conformity bias \cite{liu2016you, wang2014amazon}, position bias \cite{joachims2017accurately, joachims2007evaluating} and popularity bias \cite{abdollahpouri2020multi,wei2021model,chen2023adap,zhao2022popularity}. 
To address this issue, the academic community has probed into a multitude of methodologies to rectify the bias in recommendation systems. Given the focus of this study on selection bias, we primarily concentrate our review on the latest advancements in tackling this particular bias. For a more comprehensive understanding, we recommend readers to refer to the bird's-eye-view survey \cite{chen2023bias} for additional details.

Recent work on selection bias can be mainly categorized into three types: 

1) Generative Models, which resorts to a causal graph to depict the generative process of observed data and infer user true preference accordingly. The most representative methods are \cite{chen2018social,marlin2009collaborative,kim2014bayesian,hernandez2014probabilistic}, which jointly model the which rating value the user gives and which items
the user select to rate. More recently, some researchers utilize the causal graph to disentangle the invariant preference from other variant factors \cite{wang2022invariant, wang2020information} thereby enabling the recommendation to depend on the reliable invariant user preferences. 

2) Inverse Propensity Score, which adjusts the data distribution by reweighing the observed samples with the inverse of the propensity. Once the propensity reaches the ideal value, IPS could provide an unbiased estimation of the ideal loss. Recent studies \cite{schnabel2016recommendations, wang2021combating} have introduced a range of methodologies to learn propensities including calculating from item popularity, fitting a model to the observation, or computing from a limited set of unbiased data. 

3) Doubly Robust Learning, which enhances IPS by incorporating error imputation for all user-item pairs. DR enjoys the doubly robust property where unbiasedness is guaranteed if either the
imputed values or propensity scores are accurate. The merit of DR relies on the accuracy of the imputation model. Thus, various learning strategies are proposed by recent work. For example, DR-JL \cite{wang2019doubly} jointly learn the recommendation model and imputation model from the observed data, while the imputation model is optimized to minimize the error deviation on observed data; AutoDebias \cite{chen2021autodebias} leverages the unbiased data to supervise the learning of the imputation via meta-learning; MRDR \cite{guo2021enhanced} considers the variance reduction in learning imputation model; DR-BIAS \cite{dai2022generalized} learns the imputation with
balancing the variance and bias. More recently, some  researchers consider to further boost the instability and generalization of DR with leveraging the stable regularizer \cite{li2023stabledr} and non-parameter imputation module \cite{li2023tdr}. While these approaches offer promising solutions for debiasing recommendation, they all impute the error for all user-item pairs and may suffer from the issue of poisonous imputation.

% For example, \cite{schnabel_recommendations_2016} took recommendation as a form of treatment and introduced the inverse propensity score(IPS) and self-normalized IPS(SNIPS) methods to address bias in explicit feedback data. \cite{saito_unbiased_2020} extended this approach to implicit recommendation. \cite{wang_doubly_2019} proposed a doubly robust method by combining EIB estimator and IPS estimator. MRDR \cite{guo_enhanced_2021} method further reduce the bias and variance of the DR-JL estimator. Recently, several methods based on DR-JL were proposed, such as Multi-task DR-JL \cite{zhang_large-scale_2020},AutoDebias \cite{chen_autodebias_2021}, dr-bias \cite{dai_generalized_2022},  TDR\cite{li_tdr-cl_2023} and stableDR \cite{li_stabledr_2023}. In addition, a number of prior works \cite{bonner_causal_2018, dugang_general_2020, wang_combating_2021, wang_invariant_2022} have developed debiasing models that utilize the limited pool of unbiased data that is available. While these approaches offer promising solutions for debiasing recommendations, they have not yet addressed the issue of 'poisonous imputation' in DR-JL. Our proposed CDR method provides a more effective means of reducing the incidence of 'poisonous imputation' and is theoretically supported. 

\textbf{Uncertainty Estimation.} Utilization of probabilistic models to assess and control uncertainty (\aka variance), has found broad applications across numerous fields. This approach is usually characterized by probabilistic inference, which allows for continuous updating of beliefs about model parameters. Uncertainty estimation have found extensive use in diverse domains including machine learning \cite{loquercio2020general}, natural language processing, signal processing and clustering \cite{zhou2022comprehensive}. A prevalent approach incorporates Bayesian neural networks \cite{blundell2015weight, kingma2015variational, molchanov2017variational, louizos2017multiplicative}, providing a flexible and efficient framework to encapsulate uncertainty within neural network predictions. Another line for uncertainty estimation is the MC-dropout technique \cite{gal2016dropout, gal2017concrete, mcbook}, which simply perform multiple dropout and estimate the uncertainty (variance) via different models after dropout. Recent work has connected MC-dropout with Bayesian inference and shows that MC-dropout serves as a form of variational Bayesian inference with leveraging a spike and slab variational distribution. Besides, methods like Kronecker Factored Approximation (KFAC) \cite{ritter2018scalable} and Markov Chain Monte Carlo (MCMC) \cite{mandt2017stochastic, 2011bayesian} have been deployed to propagate uncertainties in intricate models. In this work, we simply choose MC-dropout to estimate the uncertainty of the imputation model, while it can be easily replaced by other advanced technologies. 

% Bayesian uncertainty has explored the use of probabilistic models to quantify and manage uncertainty in a variety of applications. Bayesian uncertainty methods are characterized by their probabilistic approach to infer and update beliefs about model parameters. The application of bayesian uncertainty methods has been widely adopted in fields such as machine learning, natural language processing, and signal processing, among others. One common approach is to use bayesian neural networks\cite{blundell_weight_2015,kingma_variational_2015,molchanov_variational_2017,louizos_multiplicative_2017}, which offer a flexible and effective framework to capture uncertainty in neural network predictions. MC-dropout\cite{gal_dropout_2016, gal_concrete_2017, mcbook}  employed a spike and slab variational distribution to approximate dropout at test time, thereby utilizing it as a form of variational Bayesian inference. Other methods such as Kronecker factored approximation(KFAC)\cite{ritter_scalable_2018} and Markov chain Monte Carlo(MCMC)\cite{mandt_stochastic_2017, 2011bayesian} have also been used to propagate uncertainties in complex models. Bayesian uncertainty provides a powerful tool for modeling and managing uncertainty in a wide range of applications. As far as we know, this is the first paper to introduce Bayesian uncertainty into debiasing recommendation.

\section{Conclusion and Future Work}

This study identifies the issue of poisonous imputation in recent Doubly Robust (DR) methods -- these methods indiscriminately perform imputation on all user-item pairs, including those with poisonous imputations that significantly deviate from the truth and negatively impact the debiasing performance. To counter this problem, we introduce a novel Conservative Doubly Robust (CDR) strategy that filters out poisonous imputation by examining the mean and variance of the imputation value. Both theoretical analyses and empirical experiments have been conducted to validate the superiority of our proposal. 

For future research, it would be compelling to explore more advanced filtering protocols. Our CDR strategy is based on the assumption on Gaussian distribution of the imputation, which may not be high accurate. Employing sophisticated techniques such as Dynamic Graph neural network \cite{bei2023cpdg}, Generative Adversarial Networks (GAN) \cite{creswell2018generative} or diffusion models \cite{croitoru2023diffusion} to account for more flexible distributions could be promising. Moreover, as per Table 3, DR methods typically exhibit much more computational burden compared to basic models. Therefore, investigating methods to accelerate DR presents another promising direction for future work.

\begin{acks}
This work is supported by the National Key Research and Development Program of China (2021ZD0111802), the National Natural Science Foundation of China (61972372), the Starry Night Science Fund of Zhejiang University Shanghai Institute for Advanced Study (SN-ZJU-SIAS-001) and the advanced computing resources provided by the Supercomputing Center of Hangzhou City University.

\end{acks}


\appendix
\section{Proof of Lemma \ref{la1}}
Note that the errors $e_{ui}$ and $\hat e_{ui}$ are often defined as positive values, \eg in the context of BCE loss or RMSE loss. Consequently, we can deduce $P(|{e_{ui}} - {{\hat e}_{ui}}| < {e_{ui}}) = P({{\hat e}_{ui}} - 2{e_{ui}} < 0)$. Moreover, even in cases where the positive of $e_{ui}$ and $\hat e_{ui}$ is not maintained for certain losses, we still have the relations $P(|{e_{ui}} - {{\hat e}_{ui}}| < {e_{ui}}) \ge P({{\hat e}_{ui}} - 2{e_{ui}} < 0)$. Thus, we would like to take the $P({{\hat e}_{ui}} - 2{e_{ui}} < 0)$ for analyses. 

For convenient, let $g={{\hat e}_{ui}} - 2{e_{ui}}$. Considering ${{\hat e}_{ui}}$ and ${e_{ui}}$ are two independent variables subject to gaussian distribution  $ \mathcal N(\hat \mu_{ui},\hat \sigma^2_{ui})$ and $\mathcal N(\mu_{ui},\sigma^2_{ui})$ respectively, we can easily write the distribution of $g$ as $\mathcal N(\hat \mu_{ui}-2\mu_{ui},\hat \sigma^2_{ui}+4\sigma^2_{ui})$ \cite{bishop2006pattern}. Let $z$ be a variable from standard gaussian distribution. We further have:
\begin{equation}
    \begin{split}
        \label{eq:pfla}
    P(g < 0)
       & = P(\frac{{g - ({{\hat \mu }_{ui}} - 2{\mu _{ui}})}}{{\sqrt {\hat \sigma _{ui}^2 + 4\sigma _{ui}^2} }} <  - \frac{{{{\hat \mu }_{ui}} - 2{\mu _{ui}}}}{{\sqrt {\hat \sigma _{ui}^2 + 4\sigma _{ui}^2} }})\\
       & \ge P(z <  - \frac{{{{\hat \mu }_{ui}} - 2({{\hat \mu }_{ui}} - {\varepsilon _\mu })}}{{\sqrt {\hat \sigma _{ui}^2 + 4(\hat \sigma _{ui}^2 + \varepsilon _\mu ^2)} }})\\
       & = P(z < \frac{{{{\hat \mu }_{ui}} - 2{\varepsilon _\mu }}}{{\sqrt {5\hat \sigma _{ui}^2 + 4\varepsilon _\mu ^2} }})
    \end{split} 
\end{equation}
where the inequality holds, as $\hat \mu_{ui}$, $\mu_{ui}$, $\hat \sigma_{ui}$, $\sigma_{ui}$ are bounded with $|\hat \mu_{ui}-\mu_{ui}|\leq \varepsilon_\mu$, $|\hat \sigma_{ui}^2-\sigma_{ui}^2|\leq \varepsilon_\sigma^2$, $2\varepsilon_\mu \leq \hat \mu_{ui}$. And when ${\mu _{ui}} = {{\hat \mu }_{ui}} - {\varepsilon _\mu },\sigma _{ui}^2 = \hat \sigma _{ui}^2 + \varepsilon _\mu ^2$, the right-hand side achieves minimum. Eq.(\ref{eq:pfla}) further has the following lower bound:
\begin{equation}
    \begin{split}
    &P(z < \frac{{{{\hat \mu }_{ui}} - 2{\varepsilon _\mu }}}{{\sqrt {5\hat \sigma _{ui}^2 + 4\varepsilon _\sigma ^2} }})\\ & \mathop   \ge \limits_1 P(z < \frac{{{{\hat \mu }_{ui}} - 2{\varepsilon _\mu }}}{{\sqrt 5 {{\hat \sigma }_{ui}} + 2{\varepsilon _\sigma }}})\\
     &= P(z < \frac{{{{\hat \mu }_{ui}}}}{{\sqrt 5 {{\hat \sigma }_{ui}}}} - (\frac{{2{{\hat \mu }_{ui}}{\varepsilon _\sigma }}}{{\sqrt 5 {{\hat \sigma }_{ui}}(\sqrt 5 {{\hat \sigma }_{ui}} + 2{\varepsilon _\sigma })}} + \frac{{2{\varepsilon _\mu }}}{{\sqrt 5 {{\hat \sigma }_{ui}} + 2{\varepsilon _\sigma }}}))\\
    &\mathop  \ge \limits_2 P(z < \frac{{{{\hat \mu }_{ui}}}}{{\sqrt 5 {{\hat \sigma }_{ui}}}} - (\frac{{2{M_\mu }{\varepsilon _\sigma }}}{{{\sqrt 5 m_\sigma }(\sqrt 5 {m_\sigma } + 2{\varepsilon _\sigma })}} + \frac{{2{\varepsilon _\mu }}}{{\sqrt 5 {m_\sigma } + 2{\varepsilon _\sigma }}}))
\end{split} 
\end{equation}
where the first inequaility holds due to the fact that $\sqrt 5 {{\hat \sigma }_{ui}} + 2{\varepsilon _\sigma } \ge \sqrt {5\hat \sigma _{ui}^2 + 4\varepsilon _\sigma ^2}$, while the second inequaility holds since ${{{\hat \mu }_{ui}}}$ is upper-bounded by ${{M_\mu }}$ and ${{{\hat \sigma }_{ui}}}$ is lower-bounded by ${{m_\sigma }}$.

If we let:
\begin{equation}
    \frac{{{{\hat \sigma }_{ui}}}}{{{{\hat \mu }_{ui}}}} < {\bigg (\sqrt 5 {\Phi ^{ - 1}}(\rho ) + \frac{{2{M_\mu }{\varepsilon _\sigma }}}{{{m_\sigma }(\sqrt 5 {m_\sigma } + 2{\varepsilon _\sigma })}} + \frac{{2\sqrt 5 {\varepsilon _\mu }}}{{\sqrt 5 {m_\sigma } + 2{\varepsilon _\sigma }}}\bigg )^{ - 1}}
\end{equation}
We can find the following inequaility holds:
\begin{equation}
P(z < \frac{{{{\hat \mu }_{ui}}}}{{\sqrt 5 {{\hat \sigma }_{ui}}}} - (\frac{{2{M_\mu }{\varepsilon _\sigma }}}{{{m_\sigma }(\sqrt 5 {m_\sigma } + 2{\varepsilon _\sigma })}} + \frac{{2{\varepsilon _\mu }}}{{\sqrt 5 {m_\sigma } + 2{\varepsilon _\sigma }}})) \geq \rho
\end{equation}
Thus, we have $\mathbb P(|\hat e_{ui}-e_{ui}|<e_{ui}) \geq \rho$. The lemma gets proof.

\section{Proof of Lemma \ref{la2}}
The bias and variance of CDR can be easily obtained based on the following equations:
\begin{equation}
    \begin{split}
    Bias[\mathcal{L}_{CDR}] &=\mathbb |E_o[\mathcal{L}_{CDR}]-\mathcal{L}_{Ideal}| \\ &= \frac{1}{|{\mathcal D}|} \lvert \! \! \! \sum\limits_{(u,i) \in {\mathcal D}}\! \! \! {\frac{{({p_{ui}}\! - \!{{\hat p}_{ui}})}}{{{{\hat p}_{ui}}}}({\gamma _{ui}}} ({e_{ui}} \! -\!  {{\hat e}_{ui}}) \! + \! (1\!  -\!  {\gamma _{ui}}){e_{ui}})\rvert \\
    Var[\mathcal{L}_{CDR}] & = \mathbb E_o[(\mathcal{L}_{CDR}-\mathbb E_o[\mathcal{L}_{CDR}])^2] \\ &=\frac{1}{|{\mathcal D}|^2}\! \! \! \sum\limits_{(u,i) \in {\mathcal D}} \! \! \! {\frac{{{p_{ui}}(1 \!-\! {p_{ui}})}}{{\hat p_{ui}^2}}} ({\gamma _{ui}}{({{\hat e}_{ui}}\! -\! {e_{ui}})^2}\! + \!(1\! - \!{\gamma _{ui}})e_{ui}^2)
    \end{split}
\end{equation}

The proof of tail bound refers to \cite{wang2019doubly} but replaces the $\mathcal L_{DR}$ with $\mathcal L_{CDR}$. We first let  ${l_{ui}} = \frac{{{o_{ui}}{e_{ui}}}}{{{{\hat p}_{ui}}}} + {\gamma _{ui}}{{\hat e}_{ui}}(1 - \frac{1}{{{{\hat p}_{ui}}}})$. Note that ${{o_{ui}}}$ is an bernoulli variable and thus the variable $l_{ui}$ takes the value in the interval $[{\gamma _{ui}}{{\hat e}_{ui}},\frac{{{e_{ui}}}}{{{{\hat p}_{ui}}}} + {\gamma _{ui}}{{\hat e}_{ui}}(1 - \frac{1}{{{{\hat p}_{ui}}}})]$ of size $s_{ui}=(1 - {\gamma _{ui}})\frac{{{e_{ui}}}}{{{{\hat p}_{ui}}}} + {\gamma _{ui}}\frac{{{e_{ui}} - {{\hat e}_{ui}}}}{{{{\hat p}_{ui}}}}$. Considering the $o_{ui}$ are independent for different $(u,i)$, Hoeffding inequality \cite{mohri2018foundations} can be employed with:
\begin{equation}
P(|\sum\limits_{u,i} {{l_{u,i}}}  - {\mathbb E_o}[ {\sum\limits_{u,i} {{l_{ui}}} }]|\geq |\,\mathcal D|\epsilon)\leq 2\exp({\frac{{ - 2|\mathcal D|^2\epsilon^2}}{{\sum\limits_{u,i} {s_{ui}^2} }}})
\end{equation}
Set the right-hand side of the inequality to $\kappa$ and then we can get the lemma 2.

\section{Proof of Corollary \ref{co}}
Here we primarily concentrate on demonstrating that CDR outperforms IPS in terms of variance and tail bound. A similar proof process can be applied to DR. Setting $\rho_0=0.6$ allows us to derive a set of effective imputations $S = \{ (u,i)|P(|{e_{ui}} - {{\hat e}_{ui}}| < {e_{ui}}) \geq \rho_0 \}$. If $S = \emptyset$, then CDR regresses to IPS, at least performing equivalently to IPS. Otherwise, it is always possible to identify a user-item pair $(u^*,i^*)$ that has the largest $\frac{{{{\hat \mu }_{ui}}}}{{ {{\hat \sigma }_{ui}}}}$ among $S$. We can define $\rho=\Phi (\frac{{{{\hat \mu }_{ui}}}}{{\sqrt 5 {{\hat \sigma }_{ui}}}} - (\frac{{2{M_\mu }{\varepsilon _\sigma }}}{{{\sqrt 5 m_\sigma }(\sqrt 5 {m_\sigma } + 2{\varepsilon _\sigma })}} + \frac{{2{\varepsilon _\mu }}}{{\sqrt 5 {m_\sigma } + 2{\varepsilon _\sigma }}}))-eps$, under the condition that only the imputation with the highest $\frac{{{{\hat \mu }_{ui}}}}{{{{\hat \sigma }_{ui}}}}$  is preserved, where $eps$ denotes a sufficiently small positive value. Taking into account the continuous values of ${{{{\hat \sigma }_{ui}}}},{{{{\hat \mu }_{ui}}}}$, the probability of two imputations sharing the exact same value is negligible. Hence, only the imputation for the pair $(u^*,i^*)$ is preserved.

To compare the variance and tail bounds between CDR and IPS, we can identify that the key difference pertains to the pair $(u^*,i^*)$. Here CDR utilizes $(\hat e_{ui}-e_{ui})^2$ while IPS utilizes $e_{ui}^2$. As the relation $|{e_{ui}} - {{\hat e}_{ui}}| < {e_{ui}}$ holds for $(u^*,i^*)$ with at least $\rho$ probability, and considering $\rho>\rho_0$, we can conclude that CDR achieves better variance and tail bound compared to DR. 

\bibliographystyle{ACM-Reference-Format}
\bibliography{CDR}

\end{document}